\documentclass{emulateapj}
\usepackage{graphicx}

\newcommand{\psr}{PSR~J1614$-$2230}

\begin{document}

\title{The Massive Pulsar PSR~J1614$-$2230: Linking 
Quantum Chromodynamics, Gamma-ray Bursts, and Gravitational Wave
Astronomy}

\author{Feryal \"Ozel$^1$, Dimitrios Psaltis$^1$, Scott Ransom$^2$, 
Paul Demorest$^2$, Mark Alford$^3$}
\affil{$^1$Department of Astronomy, University of Arizona, 933 N. 
Cherry Ave., Tucson, AZ 85721}
\affil{$^2$ National Radio Astronomy Observatory, 520 Edgemont Rd., 
Charlottesville, VA 29903}
\affil{$^3$ Physics Department, Washington University, St. Louis, 
MO 63130}

\begin{abstract}
The recent measurement of the Shapiro delay in the radio pulsar \psr\
yielded a mass of 1.97$\pm$0.04~$M_\odot$, making it the most massive
pulsar known to date. Its mass is high enough that, even without an
accompanying measurement of the stellar radius, it has a strong impact
on our understanding of nuclear matter, gamma-ray bursts, and the
generation of gravitational waves from coalescing neutron stars.  This
single high mass value indicates that a transition to quark matter in
neutron-star cores can occur at densities comparable to the nuclear
saturation density only if the quarks are strongly interacting and are
color superconducting.  We further show that a high maximum
neutron-star mass is required if short duration gamma-ray bursts are
powered by coalescing neutron stars and, therefore, this mechanism
becomes viable in the light of the recent measurement. Finally, we
argue that the low-frequency ($\le 500$~Hz) gravitational waves
emitted during the final stages of neutron-star coalescence encode the
properties of the equation of state because neutron stars consistent
with this measurement cannot be centrally condensed. This will
facilitate the measurement of the neutron star equation of state with
Advanced LIGO/Virgo.
\end{abstract}


\section{Introduction}

Neutron stars are associated with the most diverse and energetic
phenomena in the Universe, from gamma-ray bursts to the emission of
gravitational waves and from periodic millisecond radio signals to
month-long X-ray outbursts. The strength and even the occurrence of
some of these phenomena depend on the neutron star mass. For decades,
precise dynamical measurements of the masses of neutron stars resulted
in highly clustered values around 1.25--1.4 solar
masses~\citep{tc99}. This paradigm has recently changed with the
measurement of the neutron star mass for \psr.

\psr\ is a 3.1~ms radio pulsar in a 8.7~day orbit around a massive white 
dwarf~\citep{hess05}. Its very high inclination, at 89.17$^\circ$,
allowed the detection of a strong Shapiro delay signature in highly
accurate timing observations with the Green Bank Telescope and a
precise measurement of the neutron star mass. The inferred value of
1.97$\pm 0.04~M_\odot$ is by far the highest observed from any neutron
star to date~\citep{detal10}.

The high mass of \psr\ provides a lower limit on the maximum mass of
neutron stars and sets the dividing line between neutron stars and
black holes to at least this value. This has significant implications
for the fraction of neutron stars that will collapse into black holes
as a result of mass accretion in X-ray binaries. For a maximum neutron
star mass that is $ > 2~M_\odot$, the relative frequency of black hole
binaries that are formed through this evolutionary channel is at most
$25\%$ (Pfahl, Rappaport, \& Podsiadlowski 2003). Such a small
fraction would help explain the lack of low-mass black holes in binary
systems with dynamical mass measurements~\citep{ozeletal10}.

The highest neutron star mass that can be supported against collapse
depends very sensitively on the underlying equation of state. In
particular, if quark, hyperon, or boson degrees of freedom are excited
at high densities, the equation of state softens and cannot support
massive neutron stars (see, e.g., Lattimer \& Prakash 2001). The
discovery of even a single very high mass neutron star can, therefore,
provide strong constraints on the fundamental properties of ultradense
matter.  In this {\em Letter\/}, we explore in detail the constraints
imposed by the recent mass measurement on the radii of neutron stars
and on the properties of quark matter in their interiors.

The maximum neutron star mass also plays a crucial role in determining
the outcome of the coalescence of two neutron stars. These
astrophysical events are thought to be the central engines of
short-duration gamma-ray bursts (see reviews by Nakar 2007 and Lee \&
Ramirez-Ruiz 2007). Furthermore, they are expected to be the primary
sources of gravitational waves that will be detected by ground-based
observatories such as LIGO, Virgo, and GEO600. In \S3, we investigate
the viability of the coalescing neutron-star scenario for
short-duration gamma-ray bursts in view of the above lower limit on
the maximum mass of neutron stars. Finally, in \S4, we discuss the
prospect of measuring the equation of state of neutron-star matter
using ground-based gravitational wave observatories.

\section{The Structure and Composition of Neutron Stars}

The high mass of the \psr, in conjuction with the requirement of
causality for the neutron star matter equation of state, places a
strong lower limit on its radius, independent of the composition of
its interior. Assuming an equation of state that arises from normal
interactions in the outermost neutron star layers (where the density
$\rho$ is less than a characteristic value $\rho_0$) and requiring
that the sound speed $c_s$ everywhere in the neutron star core is less
than the speed of light $c$, such that $dP / d\rho < c^2$, leads to a
firm lower limit on the radius $R$ of a neutron star as a function of
its mass $M$~\citep{lind84,glen00,ksf97}. This is, of course, true only
for neutron stars, which are gravitationally bound and have densities
that drop smoothly to zero in their atmospheres; the causality limit
does not apply to strange stars.

The BPS equation of state (Baym, Pethick, \& Sutherland 1971) is a
standard choice for low-density matter. Assuming it is accurate up to
a density $\rho_0 = 3 \times 10^{14}$~g~cm$^{-3}$, which is comparable
to the nuclear saturation density of atomic nuclei $\rho_{\rm sat} =
2.7 \times 10^{14}$~g~cm$^{-3}$, the bound imposed by causality is
\begin{equation}
R \ge 2.83 \; \frac{GM}{c^2} = 8.3 \left(\frac{M}{1.97
M_\odot}\right)~{\rm km}.
\end{equation}
All realistic equations of state for gravitationally bound stars
predict radii that are either very weakly dependent on or decreasing
with increasing neutron-star mass~\citep{lp01}. The causality bound, in
conjunction with the recent spectroscopic measurements of neutron-star
radii in X-ray binaries that place them at values $\lesssim 12$~km
(\"Ozel, Baym, \& G\"uver 2010, Steiner, Lattimer, \& Brown 2010),
lead to the very narrow range $8.3~{\rm km}
\le R \le 12$~km for all neutron stars.

\begin{figure}
\centering
\includegraphics[scale=0.45]{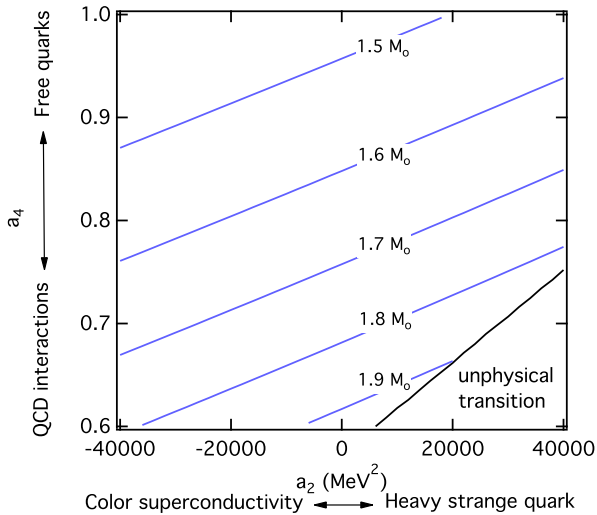}
\caption{The maximum neutron star mass as a function of two parameters 
of quark matter when the density at which the transition from
nucleonic to quark matter occurs is equal to 1.5 times the nuclear
saturation density. The measurement of a pulsar mass of $\ge
1.93~M_\odot$ from Shapiro delay observations indicates that, if the
transition to quark matter occurs at densities that are relevant to
neutron star interiors, such a massive star can be supported against
collapse only if the quarks are strongly interacting ($a_4 \le
0.63$).}
\mbox{}
\label{fig:fig1}
\end{figure}

Stronger constraints on the composition of the neutron star interior
can be placed by requiring that the equation of state support a
maximum neutron star mass that is at least as high as the measured
value. In general, the appearance of new degrees of freedom at and 
above the nuclear saturation density, such as quarks, hyperons, or 
bosons, softens the equation of state and lowers the maximum mass 
that can be supported against collapse. 

In Figure~1, we show how the mass measurement for \psr\ constrains the
quark matter equation of state. We use a specific model of quark
matter, the phenomenological equation of state proposed by Alford et
al.\ (2005), which has three parameters: $a_4$, $a_2$, and the bag
coefficient. The quartic coefficient $a_4$ measures the degree of
interaction between quarks, with $a_4=1$ corresponding to free quark
plasma. The quadratic coefficient $a_2$ depends on the mass of the
strange quark and the color superconducting energy gap; it is large in
unpaired quark matter with a heavy strange quark, and small in color
superconducting (for example color-flavor locked) quark matter with a
light strange quark. We calculate the maximum neutron star mass
$M_{\rm max}$ for a range of values of $a_4$ and $a_2$, while fixing
the bag coefficient so that the transition from nuclear matter to
quark matter occurs when the baryon density $n$ of the nuclear matter
is $1.5\,n_{\rm sat}$, where nuclear density is $n_{\rm
sat}=0.16\,{\rm fm}^{-3}$. For the nuclear matter equation of state,
we use BPS (Baym, Pethick, \& Sutherland 1971) at low densities and
APR (Akmal, Pandharipande, \& Ravenhall 1998), which we assume is
valid up to a density $2\,n_{\rm sat}$. We chose the APR equation of
state because it describes well the nucleon-nucleon scattering data
and is believed to be accurate up to the nuclear saturation
density. The excluded region in the lower right part of the figure is
the region where, after the transition from nuclear to quark matter at
$n=1.5n_{\rm sat}$, there is another transition back to nuclear matter
at $1.5 n_{\rm sat}<n<2 n_{\rm sat}$. Such a transition seems unlikely
to be physical, so we do not calculate masses in this region.

We see from Figure~1 that values of $M_{\rm max}$ compatible with the
measurement for \psr\ can only be achieved if $a_4\lesssim 0.63$ and
$a_2\lesssim 10^4\,{\rm MeV}^2$. In Alford et al.\ (2005), such values
were associated with strong interactions between the quarks (pushing
$a_4$ well below unity) and color superconductivity (which allows
$a_2$ to remain small by cancelling the contribution to $a_2$ from the
strange quark mass).

\begin{figure*} 
\centering
\includegraphics[scale=0.45]{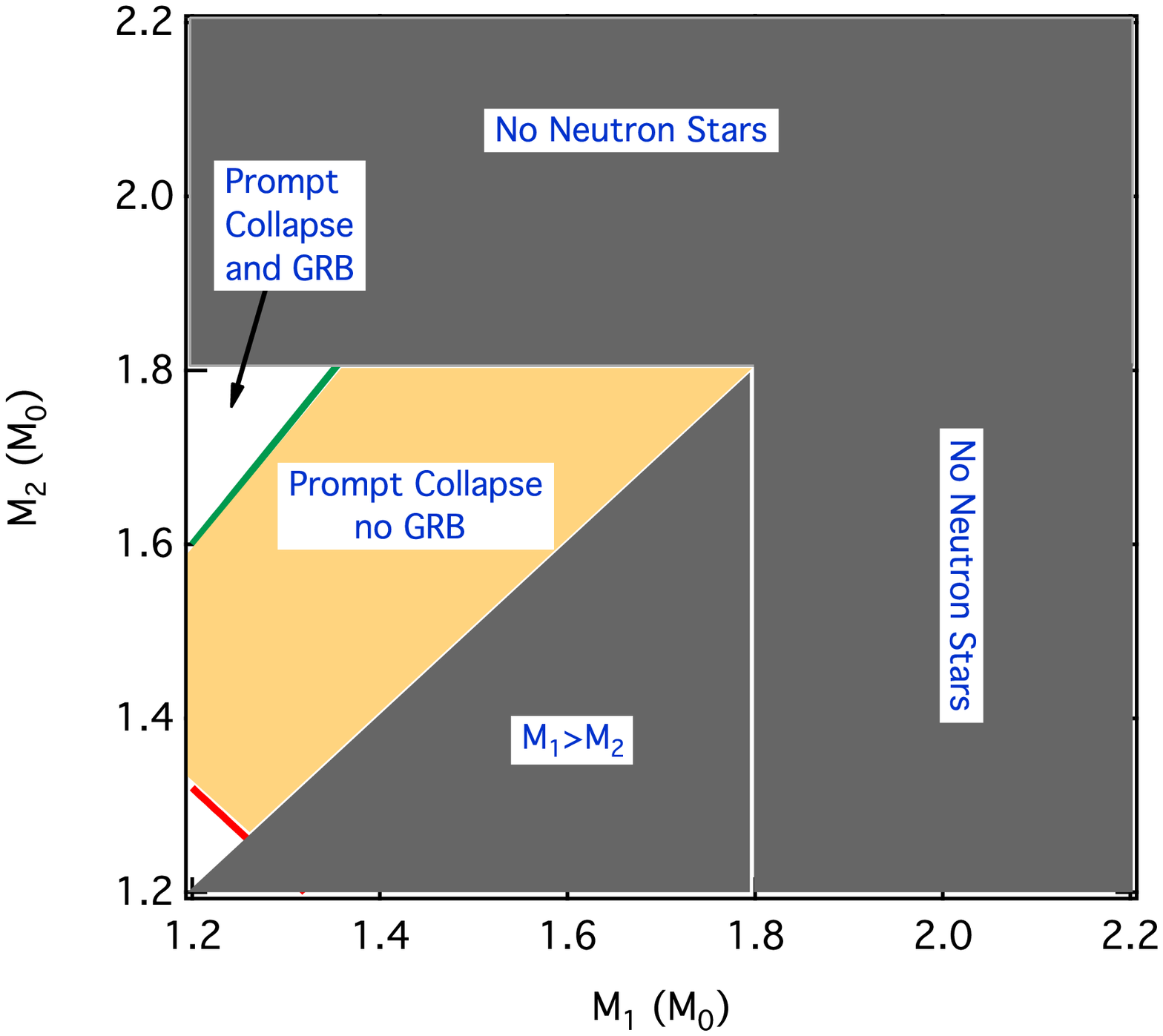}
\includegraphics[scale=0.45]{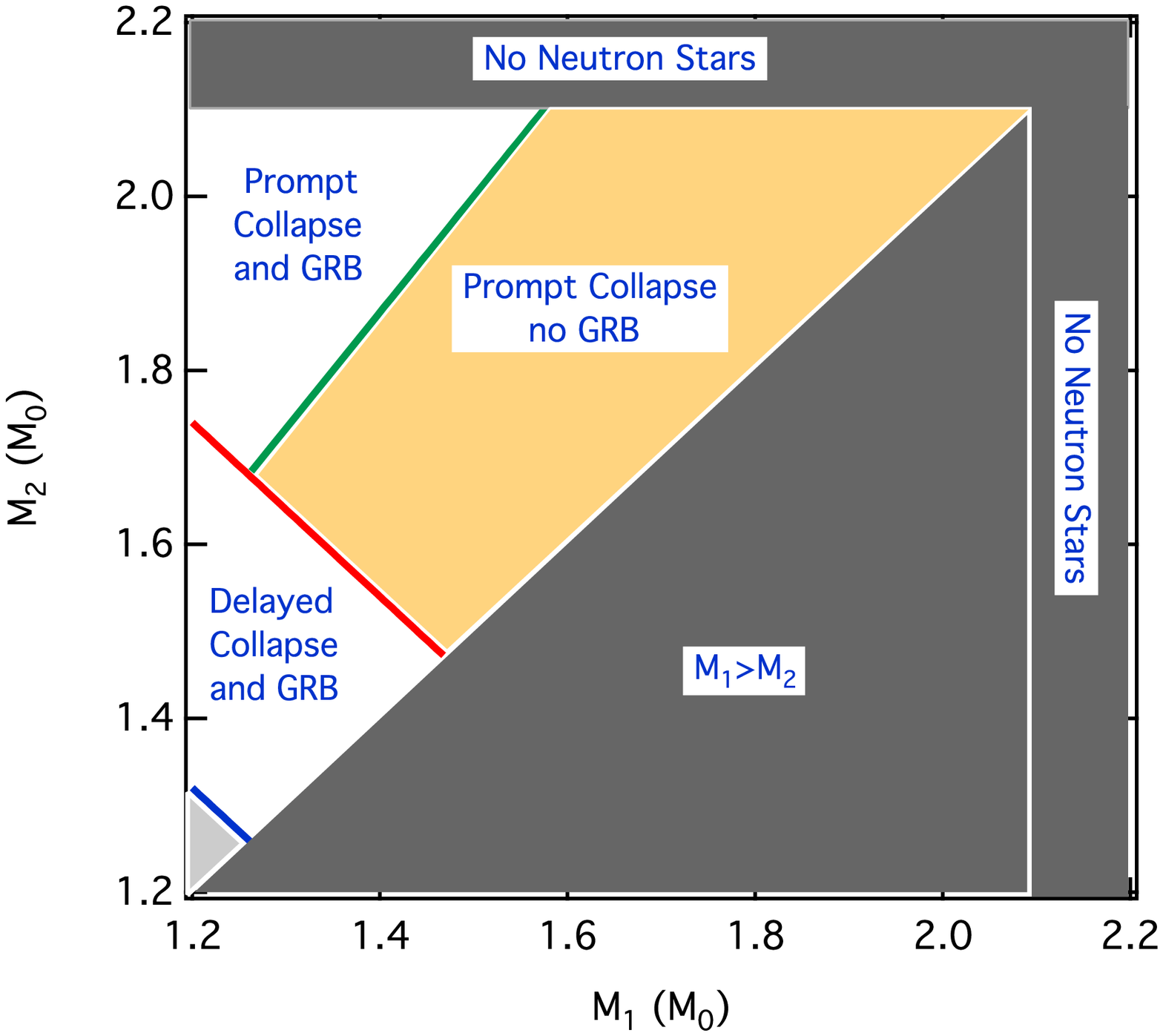}
\caption{Allowed regions of the parameter space of the masses of 
two coalescing neutron stars that leads to a short-duration gamma-ray
burst. For simplicity, the second neutron star is assumed to be larger
than the first neutron star. The left panel shows the result when the
maximum mass of a non-spinning neutron star is equal to $M_{\rm
max}=1.8~M_\odot$ and the right panel shows the result for a maximum
mass of 2.1~$M_\odot$. In the right panel, the blue line corresponds
to a total mass of $M_1+M_2=1.2~M_{\rm max}$, below which a neutron
star rotating as a solid body can be supported against collapse by
centrifugal forces. In both panels, the red line corresponds to a
total mass of $M_1+M_2=1.4~M_{\rm max}$, above which a neutron star
cannot be supported against collapse, even if it is rotating
differentially.  The green line corresponds to a mass ratio of
$M_2/M_1=4/3$, above which the initial outcome of the collision is a
black hole surrounded by a massive torus. Short duration gamma ray
bursts can be generated when the outcome of the collision is either
the delayed collapse of a supermassive neutron star into a black hole,
or the prompt collapse of the two stars into a black hole surrounded
by a massive torus. The allowed region of the parameter space for
$M_{\rm max} = 1.8~M_\odot$ is marginal but increases rapidly as
$M_{\rm max}$ exceeds 2~$M_\odot$.  }
\label{fig:fig2} 
\end{figure*}

\section{The Maximum Mass of Neutron Stars and Short-duration 
Gamma-ray Bursts}

Short gamma-ray bursts have characteristic timescales $\le 2$~s, which
are much longer than the dynamical timescale for the coalescence of
two neutron stars. In the coalescing neutron star scenario, the
required longer timescales can be achieved in two
ways~\citep{naka07,lr07,duez2010,rezz10}. The initial remnant may be a
differentially rotating massive neutron star that is temporarily
supported by centrifugal forces. The dissipation of the differential
rotation on timescales $\sim 1$~s leads to the delayed collapse of the
remnant into a black hole and produces the gamma-ray burst. In order
for this channel to occur, the sum of the masses in the initial system
has to be larger than the maximum neutron star mass for solid body
rotation but less than the maximum mass that can be supported by
differential rotation. Numerical simulations of spinning neutron stars
place these limits at $\sim 1.2~M_{\rm max}$ and $\sim 1.4~M_{\rm
max}$, respectively, where $M_{\rm max}$ is the maximum mass of a
non-spinning neutron star~\citep{duez2010,cook,bss00}; the exact value
of this limit depends on the radial profile of the differential
rotation. As a result, the delayed collapse channel requires that
\begin{equation}
1.2~M_{\rm max} \le M_1 + M_2 \le 1.4~M_{\rm max}.
\end{equation}

The second channel involves the prompt collapse of the remnant of the
coalescence into a black hole and the formation of a $\gtrsim
0.1~M_\odot$ torus around it that is accreted onto the black hole in
the required timescale. All numerical simulations of coalescing
neutron stars show that equal mass mergers hamper the formation of
these massive thick accretion disks. If, on the other hand, the ratio
of the merging neutron star masses is larger than $4:3$, the lower
mass neutron star fills its Roche lobe before contact and loses the
required amount of mass into a torus before the remnant collapses into
a black hole~\citep{st06,duez2010}. (The value of this ratio depends on
the mass of the torus required to achieve the timescale of the GRBs by
the subsequent accretion of this torus; see Rezzolla et al.\
2010). This prompt collapse channel, therefore, requires that the
total mass of the coalescing neutron stars is $\gtrsim 1.4~M_{\rm
max}$ and their mass ratio is $\gtrsim 4/3$.

In Figure~2, we calculate the outcome of mergers in the parameter
space defined by the masses of the two coalescing neutron stars and
delineate the areas in which short duration gamma-ray bursts can be
generated via either channel. The two panels demonstrate the very
strong sensitivity of these outcomes on the maximum mass that can be
supported against collapse in a non-rotating neutron star. For a
maximum neutron star mass of $1.8~M_\odot$, the allowed area in this
parameter space is marginal (left panel) but increases rapidly as
$M_{\rm max}$ grows to values larger than $2~M_\odot$ (right
panel). Therefore, the unequivocal detection of a 1.97~$M_\odot$
neutron star makes the idea of coalescing neutron stars as the central
engines of short duration gamma-ray bursts viable. 

Figure~2 shows that the largest part of the parameter space that leads
to short duration gamma-ray bursts is populated by double neutron star
systems in which one of the neutron stars is significantly more
massive than the other. This is in contrast to the double neutron star
systems observed in the Galactic disk, where both neutron stars have
surprisingly similar masses~\citep{tc99}. Indeed, a significant amount
of mass transfer onto one of the two neutron stars is likely to take
place only in a low-mass X-ray binary system, which does not lead to
the formation of a double neutron star. Double neutron stars with high
mass ratios are perhaps most easily formed in globular clusters via
exchange interactions. Globular clusters have large populations of
low-mass X-ray binaries and, therefore, may harbor the systems
responsible for the short duration gamma-ray bursts (see also Grindlay
et al.\ 2007). This would explain the recent observations of optical
counterparts of short duration gamma-ray bursts, which appear to lie
outside the host galaxies (e.g., Berger 2010).

\section{Measuring the Equation of State of Neutron Stars with 
Gravitational Waves}

The coalescence of two neutron stars is also expected to be the
primary source of gravitational waves detected with current and future
ground-based detectors such as LIGO, VIRGO, and
GEO~600~\citep{ande}. Detailed modeling of the waveforms during the
final stages of the coalescence can place strong constraints on the
equation of state of neutron-star matter~\citep{read}. This will be
possible with a generation of detectors beyond Advanced LIGO, as it
requires following the gravitational wave signal to frequencies $\ge
1$~kHz. However, even the low-frequency waveforms may encode the
characteristics of the equation of state as they are affected by the
tidal deformation experienced by the two neutron stars at larger
separations~\citep{fh08}.

The degree of tidal deformation, as measured by the Love number, is
relatively insensitive to the compactness of the neutron star but
depends strongly on how centrally condensed the density profile inside
the stars is~\citep{hind,dn09,bp09}, because stars that are centrally
condensed are weakly deformed in the presence of an external tidal
gravitational field.

Equations of state that allow pulsar masses $\ge 2 M_\odot$ typically
yield neutron stars that are not centrally condensed (as can be
inferred from the polytropic models of Hinderer 2008). Therefore, the
discovery of the 1.97~$M_\odot$ pulsar indicates that coalescing
neutron stars will undergo significant tidal deformation, and
information about the equation of state of the stellar interior will
be encoded in low-frequency ($\le 600$~Hz) gravitational waves emitted
during coalescence. Detection of these waves will allow accurate
measurements of the equation of state of neutron-star matter in the
near future.

\acknowledgments

We thank Sanjay Reddy, Luciano Rezzola, and Chryssa Kouveliotou for
useful discussions and comments. F\"O acknowledges support from the
NASA ADAP grant NNX10AE89G, NSF grant AST 07-08640, and Chandra Theory
grant TMO-11003X. DP was supported by the NSF CAREER award NSF 0746549
and Chandra Theory grant TMO-11003X. PD is a Jansky Fellow of the
National Radio Astronomy Observatory. The National Radio Astronomy
Observatory is a facility of the National Science Foundation operated
under cooperative agreement by Associated Universities, Inc.


\clearpage

\end{document}